\documentclass[structabstract]{aa}
\usepackage{graphicx}
\usepackage{txfonts}
\usepackage{longtable}
\begin{document}
   \title{Subsurface chemistry of mantles
          of interstellar dust grains
          in dark molecular cores}

   \author{J. Kalv$\mathrm{\bar{a}}$ns
          \inst{1}
          \and
          I. Shmeld
          \inst{2}
          }

   \institute{Institute of Astronomy, University of Latvia, Raina 19, Riga, Latvia\\
              \email{kalvans@lu.lv}
         \and
             Ventspils International Radioastronomy Centre of Ventspils University College, In$\mathrm{\check{z}}$enieru iela 101, Ventspils, Latvia\\
             \email{ivarss@venta.lv}
             }

   \date{Received February 3, 2010; accepted May 13, 2010}

  \abstract
   {The abundances of many observed compounds in interstellar molecular clouds still lack an explanation, despite extensive research that includes both gas and solid (dust-grain surface) phase reactions.}
   {We aim to qualitatively prove the idea that a hydrogen-poor subsurface chemistry on interstellar grains is responsible for at least some of these chemical ``anomalies''. This chemistry develops in the icy mantles when photodissociation reactions in the mantle release free hydrogen, which escapes the mantle via diffusion. This results in serious alterations of the chemical composition of the mantle because pores in the mantle provide surfaces for reactions in the new, hydrogen-poor environment.}
   {We present a simple kinetic model, using existing astrochemical reaction databases. Gas phase, surface and subsurface pore reactions are included, as are physical transformations of molecules.}
   {Our model produces significantly higher abundances for various oxidized species than most other models. We also obtain quite good results for some individual species that have adequate reaction network. Thus, we consider that the hydrogen-poor mantle chemistry may indeed play a role in the chemical evolution of molecular clouds.}
   {The significance of outward hydrogen diffusion has to be proved by further research. A huge number of solid phase reactions between many oxidized species is essential to obtain good, quantitative modeling results for a comparison with observations. We speculate that a variety of unobservable hydrogen-poor sulfur oxoacid derivatives may be responsible for the ``disappearance'' of sulfur in dark cloud cores.}

  \maketitle
%

\section{Introduction}
\label{intro}
 It is widely accepted that molecular hydrogen and many other interstellar molecules form on  interstellar dust grains. There has been wide research in the field of gas-grain chemistry occurring in the dark, dense cores of interstellar molecular clouds, including observations and calculations. Various desorption mechanisms and chemical reactions on the surfaces of the grains are investigated.

Several molecules in the interstellar medium at least are known whose abundances are not easily explained by gas-phase and grain surface-phase chemistry. These include OCS, HCN, SO, cyanopolyynes, and others. In the warming star-formation regions there appear highly oxidized organic compounds that indicate that a different chemistry occurs in these regions or, as we believe, the ejection of heavily processed interstellar grain mantles into the gas phase.

The grain surface reactions in interstellar clouds by definition are subjected to heavy hydrogenation. The proper production of heavier and hydrogen-poor species is difficult to reproduce by calculations at least in some cases (Hasegawa \& Herbst \cite{11}, van Weeren et al. \cite{01}, Hatchell et al. \cite{35}). We present an alternative explanation of this problem by considering the possibility that within the grain mantles below the surface layer there is a hydrogen-poor, chemically active material. The aim of this article is to qualitatively answer this question through existing knowledge and means of astrochemical problem solving.

   There are only few such models that take into account more than one layer of the accreted   species. There are some serious researches done which insist that the chemistry of the species  frozen onto grains does not end with the formation of the next accreted layer. Shalabiea \& Greenberg (\cite{26}) examine photon-induced processes inside the mantle. They conclude that  photoprocessing of grain mantles is the start of the synthesis of many species. Hasegawa \& Herbst (\cite{10}) use a 3-phase model to examine the formation and composition of the inner mantle. They found that radicals like OH, $\mathrm{CH_{3}}$ etc. are absorbed into mantles in large numbers. We argue that the chemistry below the outer surface of the icy mantle has to be common and is an important path in interstellar molecule synthesis. Besides, they also note that grain surface reactions tend to overproduce hydrogenated species, which is the main problem we attempt to tackle in this work. Schutte \& Greenberg (\cite{27}) examine the possibility of molecule desorption from the grains by chemical explosions within the grain mantle. Naturally, these reactions can be expected to alter the chemical composition of the mantle itself. Freund \& Freund (\cite{38}) present a dust grain model based on the principle of solid solutions,  producing results that explain important features in the molecular cloud composition.

 We present a model of the processes on the surface and inside the ``frozen'' grain  mantles with a basic concept that chemical reactions occur on the surface of pores (or  cracks) inside the mantles. The point that makes the difference between the surface and mantle  reactions is that the mantle is not directly exposed to the ocean of hydrogen in a nebula, and thus a significantly different chemistry may develop. This gives an opportunity to present an explanation for some of the astrochemical mysteries. We do not use advanced calculation  techniques or new reactions; we evaluate the importance of hydrogen diffusion through the grain mantle on the chemical composition of the mantles in dark molecular cloud cores. Thus, the model  includes various physiochemical processes but otherwise keeps a rather conservative approach.


\section{The model}
\label{model}
\subsection {Model considerations}
	\label{consider}
   Current models, e.g. Das et al. (\cite{04}), Goldsmith et al. (\cite{07}), van Weeren et al. (\cite{01}) insist that the collapse of an interstellar cloud to densities around $10^{5} cm^{-3}$ occurs at a time around $10^{6}$ years and that molecules form simultaneously. Deposition onto grains proceeds at timescales comparable to cloud cooling and collapse, and most of the accreted matter will accumulate at late stages of the cloud evolution. Taking this into consideration we developed a steady-phase chemical model for timescales much longer than the cloud formation.
   
   In order to investigate the molecular abundances in dark cloud cores under chemical equilibrium we used a chemical kinetics model with 352 molecules. The interstellar UV radiation (molecule photodissociation and photodesorption) has been neglected, the integration time was taken to be $10^{16}s$, comparable to an interstellar cloud entire lifetime, and the cloud density $n_{H}=10^{5}cm^{-3}$. The gas temperature is taken to be 15\emph{K}, the dust grain temperature -- 10\emph{K}. We used the UDFA06 dipole (\textit{udfa06}) astrochemistry database (Woodall et al. \cite{12}) is used to provide the gas-phase chemical reaction set. We adopted elemental abundances provided by Jenkins (\cite{30}). The elements permitted are H, He, C, N, O, Na, Mg, Si, S, Fe.
   
The grain chemistry is described in terms of surface reactions between species that are located on the outer grain surface or on a surface of pores inside the grain mantle itself. It might be possible to even more adequately describe the grain mantle chemistry with the concept of a solid solution or nanoporous matter. For the grain processes the surface reaction set by Hasegawa, Herbst, Leung \cite{18} and Hasegawa \& Herbst (\cite{11}) is used (neutral molecules only). Species without gas phase \textit{udfa06} reactions were excluded from calculations, because they also lack the important photodissociation sink reactions (Sect.~\ref{CRPHdiss}). These include $\mathrm{N_{2}H}$, HOC, NaOH, etc. In order to provide a more adequate reaction set for each molecule, organic molecules are included up to $\mathrm{C_{5}}$ only in the grain surface and mantle model. This is because larger molecules tend to have less surface reactions included, especially when one takes into account the number of atoms they contain.  Notably, there is an almost complete lack of oxidation reactions for the more complex carbon species. These reactions are extremely important because our model shows that a highly oxidative environment is possible in the grain mantles.

Because of the use of a huge gas-phase reaction database and a limited solid-phase reaction database, there are many gas phase molecules that do not take part in the surface and mantle processes. Since our aim is to investigate exclusively the solid phase of interstellar molecules, we decided for the sake of completeness to leave the \textit{udfa06} as intact as possible in our model, with some 260 gas-only species. Most of the molecules with a gas phase only are not relevant to the solid phase (ions), and the remainder are expected to have negligible abundances anyway (complicated organic species). When one compares similar molecules with and without a solid phase, the difference in abundances is about one order of magnitude. The full list of results for gas-phase species is given in Table~\ref{gasconc}.

We used a chemical model integrated in a 3-phase system similar that of Hasegawa \& Herbst (\cite{10}). The three phases are gas, grain surface, and grain mantle. The grain surface consists of reactive species that accrete from gas, may be easily desorbed by several mechanisms, and are subjected to photodissociation. In this context the surface consists of the first few layers of a mantle. The mantle itself is formed by buried surface molecules. H atoms and especially $\mathrm{H_{2}}$ molecules, which are created in the mantle by photolysis can migrate away from their parent molecule, and the mantle becomes enriched with free hydrogen. This excess hydrogen is can escape to the outer surface by diffusion. Thus, with photolysis a mostly hydrogen-poor chemical environment forms below the surface.

Molecules adsorbed onto grains are divided into two layers - the surface and the mantle. We employed a model of an equilibrium state, where surface and gas phase molecules are in a dynamic equilibrium, while mantle molecules are almost permanently locked away in a frozen state. Thus, ``surface'' represents a few (rugged) top layers that are subjected to desorption and surface reactions and may be brought to the very top layer by these processes. ``Mantle'', by definition, is never exposed to the surface and is only slowly returned to the gas phase by direct ejection caused by cosmic ray hits. In order to properly describe the physical and chemical processes we avoided the division of the mantle in layers. Surface and mantle molecules are treated as solid species uniformly dispersed in volume with abundances expressed in $cm^{-3}$.

\subsection {Molecule accretion onto grains}
\label{accr}
Molecule accretion onto grains happens in an equilibrium with the various desorption processes. In our model only neutral molecules accrete. The rate coefficient ($s^{-1}$) is calculated according to Willacy \& Williams (\cite{13}) and Nejad \& Wagenblast (\cite{14}) by the formula

   \begin{equation}
   \label{accCoef}
k_{accr}= 3.2\times10^{-17}n_{H}S_{i}\left(\frac{T_{g}}{M_{i}}\right)^{1/2},
	 \end{equation}

	 where $S_{i}$ is the sticking coefficient, $M_{i}$ is molecular mass of species \textit{i} in atomic mass units,
and ${n_{H}}$ is total number of H atoms per $cm^{3}$, $T_{g}$ is gas temperature (\textit{K}).

The rate of accretion, $cm^{-3}s^{-1}$, is thus

   \begin{equation}
   \label{accRate}
R_{accr}= k_{accr} n_i,
	 \end{equation}

where $n_i$ is the gas phase abundence of species $i$.

The sticking coefficient used by previous authors (i.e. Willacy \& Williams \cite{13}, Nejad \& Wagenblast \cite{14}, Roberts et al. \cite{16}, Aikawa et al. \cite{17}, Turner \cite{19}, Willacy \& Millar \cite{25}, Brown \& Charnley \cite{28}) for heavy species is 0.1 to 1, usually around 0.33, and that for light species is 0 to 1. We took an approximate average path with $S_{i}$=0.33 for all heavy species and 0.1 for hydrogen atoms and $\mathrm{H_{2}}$ molecules.
	
	 \subsection {Thermal evaporation}
	 \label {evap}
	 The thermal evaporation rate is calculated from the evaporation times given in Hasegawa \& Herbst (\cite{11}).
	
	 \subsection {Direct cosmic-ray heating desorption}
	 \label {CRdesorp}
	 Desorption by heating, caused by Fe nuclei of cosmic rays, is calculated with the rate coefficients given by Hasegawa \& Herbst (\cite{11}).

	 \subsection {Cosmic ray induced photodesorption}
	 \label {CRPHdesorp}
Rate coefficient ($s^{-1}$) for desorption by cosmic ray induced photons for surface species is calculated by the formula adapted from Willacy, Williams (\cite{13})

   \begin{equation}
   \label{CRphdes1}
k_{crpd}=R_{ph}Y,
   \end{equation}

where the photon hit rate $R_{ph}$ for a single grain
	
   \begin{equation}
   \label{CRphdes2}
R_{ph}=F_{p}\left\langle \pi\alpha^{2}_{g}\right\rangle.
   \end{equation}

Yield \textit{Y} is taken to be 0.1,
$F_{p}$ is photon flux, taken 4875 $cm^{-2}s^{-1}$ from Roberts et al. (\cite{16}). $\left\langle \pi\alpha^{2}_{g}\right\rangle$ is the average cross section of a grain ($1.0\times10^{-10} cm^{2}$). Like for other desorption mechanisms, the desorption rate is

   \begin{equation}
   \label{CRphdes3}
R_{crpd}=k_{crpd} n_{surf,i}.
   \end{equation}

$n_{surf,i}$ is the average abundance ($cm^{-3}$) of species $i$ residing on the outer surface of a grain.

	 \subsection {Dissociation by cosmic ray induced photons in solid state}
	 \label {CRPHdiss}
Besides the surface binary reactions of Hasegawa, Herbst, Leung, (\cite{18}) and Hasegawa \& Herbst (\cite{11}), we also include a limited set of cosmic ray photon induced photoreactions on grain surfaces.

We pretend that the individual molecules on and in the amorphous mantles with mixed composition essentially keep their UV absorption properties. The gas phase reaction coefficients we obtain from \textit{udfa06}. H and $\mathrm{H_{2}}$ formed in photodissociation on the outer surface of grains are always allowed to escape into gas phase, while all other species remain on the  grain surfaces (their desorption by CR induced photons is described above in Sect. ~\ref{CRPHdesorp}). We consider it a reasonable approximation given that the surface species in the model of Hasegawa \& Herbst (\cite{10}) are probably too intensively hydrogenated, and Andersson \& van Dischoeck (\cite{07}) concludes that photodissociation of amorphous water ice mostly results in an escaping hydrogen atom.

Several sources (Shalabiea \& Greenberg \cite{26}, Andersson \& van Dischoeck \cite{07}, \"{O}berg et al. \cite{23}) indicate that photoreactions can and do occur in the layers below the surface, although they usually do not lead to desorption.

Thus, the formula for photodissociation rate coefficient is
   \begin{equation}
   \label{CRphdiss1}
k_{ph.dis}=k_{udfa}Y_{dis}.
   \end{equation}
   Generally, the quantum yield is taken to be 0.1 for any given species on grain surface $Y_{s.dis}$. We assume that within the mantle only molecules with access to a surface can be effectively dissociated, so that the products do not recombine again. We expect 1/100 of the mantle to be exposed to inner pore surfaces.  The dissociation yield for mantle molecules is

\begin{displaymath}
   \label{CRphdiss2}
Y_{mantle,dis}=0.1\times10^{-2}.
\end{displaymath}

	 The dissociation products of these inert molecules are then assumed to be reactive until they become frozen again (see Sect.~\ref{reacs}).

	 \subsection {Desorption resulting from $\mathrm{H_{2}}$ formation}
	 \label {H2fdes}	
We calculated the rate coefficient ($s^{-1})$ for desorption of surface molecules by heat released by $\mathrm{H_{2}}$ molecule formation on grains according to Roberts et al. (\cite{16})
   \begin{equation}
   \label{h2fdes1}
k_{\mathrm{Hf.des}}=\frac{\epsilon R_{\mathrm{H_2}}}{n_g},
   \end{equation}
where the abundance $n_{g}$ ($cm^{-3}$) of dust grains is
   \begin{equation}
   \label{reac3}
n_{g}=1.33\times10^{-12}\times n_{H}.
   \end{equation}
${R_{\mathrm{H_2}}}$ is the formation rate of $\mathrm{H_{2}}$ molecules and $\epsilon$ is number of molecules desorbed per act of $\mathrm{H_{2}}$ molecule formation. The species desorbed are those with binding energies $E_{b}$ below a treshold energy. According to the discussion by Willacy \& Millar (\cite{25}) and Roberts et al. (\cite{16}), the efficience of desorption by heat released by $\mathrm{H_{2}}$ formation is a subject of discussion and suggestions. Following Roberts et al. (\cite{16}) we take $\epsilon$=0.01. Like in other models, this selective desorption mechanism has a strong effect on the abundances of major species in both, gas and solid phases. We choose to diminish its importance by taking $E_{b} < 1210K$ (binding energies from Hasegawa \& Herbst \cite{11}).

	 \subsection {Binary reactions on grains}
	 \label {reacs}
For chemical processes on grain surfaces, i.e. reactions between accreted species, we use the reaction set provided by Hasegawa, Herbst \& Leung (\cite{18}) and Hasegawa \& Herbst (\cite{11}). The rate coefficient is
   \begin{equation}
   \label{reac1}
k_{ij}=\frac{\kappa_{ij} (R_{diff,i} + R_{diff,j})}{n_{g}},
   \end{equation}
   where the diffusion rate is
   \begin{equation}
   \label{reac2}
R_{diff}=N^{-1}_{s}\times t^{-1}_{hop}.
   \end{equation}
   Because the roughness and inhomogeneity of the mantle are key features of this model, we used the higher quantity $2\times10^{6}$ for number surface adsorption sites $N_{s}$ instead of $10^{6}$, which characterizes a perfect spherical grain.  $\kappa_{ij}$ is the probability for the reaction to happen upon an encounter, $t_{hop}$ is the hopping time (\textit{s}) for mobile species.

We assume that physical transformations of a fully formed grain mantle in steady conditions inside dark cores are driven by cosmic rays passing through the grain. When a heavy cosmic ray particle hits the grain, the mantle is heated, shaken, and at least locally rearranged. New surfaces appear and molecules and some radicals emerge, migrate, and react, while other species become locked in ice again. Thus the mantle (and, perhaps, the grain) is not a frozen chunk of molecules but undergoes changes in timescales comparable to cloud lifetimes. Generally, reactions occurring within the icy mantles should be of great importance because most of the frozen molecules are accumulated in the volume, not the surface. Cosmic-ray-induced alternations are generally regarded as first-order reactions. We keep this approach here. Molecules in a mantle are activated with a rate coefficient ($s^{-1}$)
   \begin{equation}
   \label{reac4}
k_{act}=t^{-1}_{cr}\times u^{-1},
   \end{equation}
where $t_{cr}$ is the average time between two successive strikes by a Fe cosmic ray nucleus (calculated $3.16\times10^{13}$ s by Hasegawa \& Herbst \cite{11}) and \textit{u} is the number of strikes needed to fully reorganize the grain mantle, i.e. to expose to an inner surface an amount of molecules equivalent to the total number of molecules per mantle. \textit{u} is dependent on the porousness and thickness of an average mantle. The thickness is estimated to be 100 monolayers by Hasegawa \& Herbst (\cite{10}) and 25 layers by Turner (\cite{08}). The former value is generally more consistent with our model. We assumed \textit{u}=200.

   The reactions of the activated (exposed to a pore surface) reactants proceed with the rate coefficient (Hasegawa, Herbst \& Leung \cite{18})
   \begin{equation}
   \label{reac5}
k_{mantle,ij}=\frac{\kappa_{ij} (R_{diff,i}+R_{diff,j})}{n_{g} n_{p}}.
   \end{equation}
There is no knowledge about pores of interstellar dust; their size and numbers should be affected by the exact conditions in the dark core. However, they are limited by the number of mantle layers and size of the grain (see above). Undoubtedly, most of the pores are too small to provide a functioning reaction surface, while there should be only a few larger pores and with a higher possibility to be connected to the outside, thus indeed becoming at least partially a gully of the outer surface. For the calculation of the diffusion rate $R_{diff}$ the number of adsorption sites per average pore $N_{s}$ in Eq.~\ref{reac2} is now taken to be $10^{3}$. We take the same number for the pores per grain, $n_{p}=10^{3}$. We consider this choice a possibly neutral estimate.

The factor that changes the reaction rate taken from Hasegawa, Herbst \& Leung (\cite{18}) and Hasegawa \& Herbst (\cite{11}) is inversely the number of isolated surfaces in a grain multiplied by the number of adsorption sites on the available surface. For outer surface (as given in Hasegawa, Herbst \& Leung \cite{18}, Eqs. 4 and 9) this factor is
\begin{displaymath}
\frac{1}{1\times10^{6}\times n_{g}}.
\end{displaymath}
For pores we take
\begin{displaymath}
\frac{1}{10^{3}\times10^{3}\times n_{g}}.
\end{displaymath}
That is, we assume that there are thousand pores in an average grain with a surface of thousand adsorption sites each.

The molecules are assumed to be in the activated state from a Fe cosmic ray hit to about the time of the next cosmic ray hit. The interval of hits is assumed to be $t_{cr}$ (\textit{s}). The rate coefficient ($s^{-1}$) is thus
   \begin{equation}
   \label{reac6}
k_{inact}=t^{-1}_{cr}.
   \end{equation}
The chosen rate coefficients ensure that at any given time instant roughly 0.5\% of mantle species are ``activated'' and are taking part in reactions on the pore surfaces. It is half the number of molecules assumed to be exposed to pore surface. The remaining half is assumed to be inactive for some reason (molecules which reside in pores too small for reactions and temporarily blocked sites). The molecular abundances (gas, surface, and mantle) produced by the model are only very slightly dependent on this percentage. It is because the chemical reactions on surface are anyway much faster than the radical production on surfaces by CR induced photons.

	 \subsection {Hydrogen diffusion through mantle}
	 \label {Hdiff}
Hydrogen diffusion from the surface to the mantle pores and from the mantle to the surface is included in our model. We calculated the diffusion rate coefficient assuming that H atoms and $\mathrm{H_{2}}$ molecules mostly reside on outer or inner surfaces. The hydrogen within the mantle lattice is only a relatively rare intermediate state. Thus
   \begin{equation}
   \label{hdiff1}
k_{H,diff}=D P_{diff} L,
   \end{equation}
where \textit{D} is the diffusion coefficient, $P_{diff}$ is the probability for diffusing species to find an appropriate surface. $L$ is the length of the diffusion path (we take it $1\times10^{-6} cm$), approximately half the thickness of the mantle. We adopt \textit{D} for H atoms $2.50\times10^{-21} cm^{2} s^{-1}$ estimated by Awad et al. (\cite{29}). For the $\mathrm{H_{2}}$ molecules we use the coefficient given by Strauss \& Chen (\cite{06}) for diffusion in hexagonal ice extrapolated to 10K ($D_{H_{2}}=5.90\times10^{-8} cm^{2}s^{-1}$). This may sound rather tentative as the diffusion in amorphous ice is expected to be significantly slower. We could not find any estimates for $D_{H_{2}}$ in amorphous ice, however, in our model the real hydrogen content is rather independent of the exact value of $D_{H_{2}}$ as long as $D_{H_{2}}>>D_{H}$ and the real $\mathrm{H_{2}}$ diffusion rate is much shorter than the photodissociation rate -- both of which are expected to be true.

We use the value $P_{diff,MS}$=0.5 for outward diffusion (assuming that on average 50\% of all diffusion directions from pores lead outwards) and $P_{diff,SM}=1\times10^{-2}$ for inward diffusion because in our model we expect only 1/100 of the mantle volume to be occupied by the larger pores, able to host enough reactants on their surfaces for reactions to occur. This means that 99\% of surface hydrogen diffusing inward returns to the surface and only 1\% reach an inner pore to reside in. These probabilities result in an eqilibrium where the outward flux of hydrogen dominates. A release of hydrogen and the formation of less hydrogenated species is observed in numerous experiments involving photolysis of hydrogenated species in vacuum, e.g. Gerakines et al. (\cite{48}), Andersson \& van Dischoeck (\cite{07}).

	 \subsection {Direct ejection by cosmic rays of mantle molecules}
	 \label {eject}
Certainly there must be a process besides the transfer of gas molecules to the surface and then to the mantle that works in the opposite direction or a complete freeze-out will result, which is not observed. We propose a simple cosmic-ray-driven mechanism which consistently returns a proportion of the inner mantle into the gas phase. These are the molecules believed to be directly, unselectively ejected by a hit of a Fe cosmic ray nucleus. The rate of ejection ($cm^{-3}s^{-1}$) is calculated by
   \begin{equation}
   \label{eject1}
R_{ej}=t^{-1}_{cr} Y_{cr} n_{g} X_{mantle,i}.
   \end{equation}
$X_{mantle,i}$ is the proportion of species \textit{i} in the grain mantle. $Y_{cr}$ is the number of mantle molecules ejected by this process in each hit. It should be a negligible number compared to the number of molecules desorbed by the cosmic-ray heating mechanism, estimated to be at least $10^{6}$ by Hasegawa \& Herbst (\cite{11}). We take $Y_{cr}$ 0.01\% of this value, that is 100 molecules per hit of cosmic-ray Fe nuclei. Gas and solid phase abundances are unaffected by the value of $Y_{cr}$ over several orders of magnitude.
The returning of mantle molecules into the gas means that gas, surface, and mantle phases are in an interconnected equilibrium in our model. Molecule abundances in both surface and mantle directly affect the abundances in gas, where the species are easily observable.

	 \subsection {Surface-to-mantle transition}
	 \label {mantleform}
We calculated the transformation of the surface species into mantle species in the local thermodynamic equilibrium with a constant rate coefficient
   \begin{equation}
   \label{mantle1}
R_{mantle}=k_{circ} n_{surf,i}.
   \end{equation}
$k_{circ}$ is the total circulation rate of a surface molecule at an equilibrium state. The molecule accretes from the gas phase, undergoes surface diffusion, reactions, desorption acts, and is eventually buried by other molecules, and finally ends up in the mantle.
We treat this coefficient as an unknown variable and use the model with the given, often assumed, input conditions (Sect.~\ref{accr} to~\ref{CRPHdesorp} and~\ref{H2fdes}, and~\ref{eject}) to choose $k_{circ}$. We took $3.7\times10^{-17} s^{-1}$ to produce a ratio 30:1 between the total abundances of all atoms in the mantle and on the surface (except hydrogen). Gas, surface, and mantle abundances are very strongly affected by the value of $k_{circ}$, consequently it has to be chosen with a precision to decimals.  H and $\mathrm{H_{2}}$ are not incorporated into the mantle by this mechanism.

It is not a physically adequate approach to calculate the rate of surface-to-mantle transition. However, with so many assumptions and poorly known values used (e.g. $S_{i}$, $Y_{cr}$, $u$ etc.) we consider it satisfactory for the aims of this paper. This approach also implies that the program might produce somewhat biased results in terms of absolute abundances ($cm^{-3}$).

	 \subsection {Desorption by chemical explosions}
	 \label {explos}
We do not include desorption by chemical explosions (Schutte, Greenberg \cite{27}) of grain mantles. In our current model the radicals released by photoprocess do not react rapidly and violently as is required by the explosion theory. They are entirely consumed by the reactions on the pore surfaces. Less than 1/100 of the mantle species are radicals in calculation results of our model, partially due to the incomplete reaction network. However, the photon flux in dark cores is significantly lower than in experiments producing explosion effects. Fully researching them requires a more sophisticated model, because the explosive reactions also directly affect the chemical composition of the icy mantle.

	 \subsection {Model credibility}
	 \label {credibl}
To focus our research on chemical processes at the local thermodynamic equilibrium we investigated a standard case of a very long-existing interstellar dark cloud core, completely isolated from the interstellar UV radiation field. The parameters regarding the various physical and chemical transformations of solid phase species (Sect.~\ref{consider} -~\ref{mantleform}) were chosen in accordance with our understanding of the grain structure and processes. These parameters are able to influence the model output (calculated fractional abundances) in a wide range.

Observational results found in the literature were used to evaluate the rate of ejection of the mantle molecules by cosmic rays (Sect.~\ref{eject}) and surface-to-mantle transition (Sect.~\ref{mantleform}). These two rates essentially determine the abundances of gas phase species. We opted for long-lived (presumably LTE) interstellar cloud cores -- dark, cold, and with metals heavily depleted on grains (Hollenbach et al. \cite{15}, Turner \cite{36}, Tafalla et al. \cite{37}).

There are several aspects or assumptions in the model that require a credibility assessment. Most important is, how close the pore surface model represents the real conditions on interstellar grain mantles. Under certain (given) input conditions the model can produce feasible results. We assume that small pores are insignificant for the mantle chemistry, but they may affect the hydrogen content within the mantle to an unknown extent. Also, we do not take into account the ability of hydrogen to react while diffusing through the solid amorphous ice. Generally these effects should lead to an increased abundance of hydrogenated species at the expense of radicals (perhaps reducing the efficiency of desorption by explosions).

A second aspect regarding the validity of the model is the yields of various photoprocesses, especially the yield of the photodissociation inside the mantle. The model works best with the assumption that below the few surface layers dissociation essentially occurs on pore surfaces only. This is certainly not true, as dissociation basically is not a surface-related process. However, the assumption is justified by the higher possibility of recombination of dissociation products unable to diffuse away in a (amorphous) lattice. However, some the hydrogen released within the lattice may be able to diffuse, again increasing the radical content and supplying atomic hydrogen for reactions and diffusion to the surface. Si, Na, Mg and Fe have a very limited or nonexistent set of solid phase reactions and species and we do not include these elements in our discussion about results. Silicon is kept in the surface-mantle model in order to increase the total number of available reactions and thus improve the general accuracy of results for other species.

	 \section {Results}
	 \label {results}
We recall here that the simplified approach in the model is a quick solution to promote the discussion about the importance of hydrogen diffusion effects on grain mantle chemical composition. We emphasize that an important feature is the limited number of reactions available for each species in the mantle. A typical solid heavy molecule, for example, $\mathrm{C_{5}H_{2}}$ or SO, has one or two dissociation reactions by cosmic ray photons. Also, there are usually two to three reactions where the species is a product. There are usually no more than six reactions (production and destruction) involving a complicated molecule. It is a poor set of chemical transformations, especially in the mantles where hydrogenation reactions play a role of relatively little importance. Thus the calculated abundances for several important species are highly biased, as noted below.

\begin{figure}
	\centering
    \includegraphics{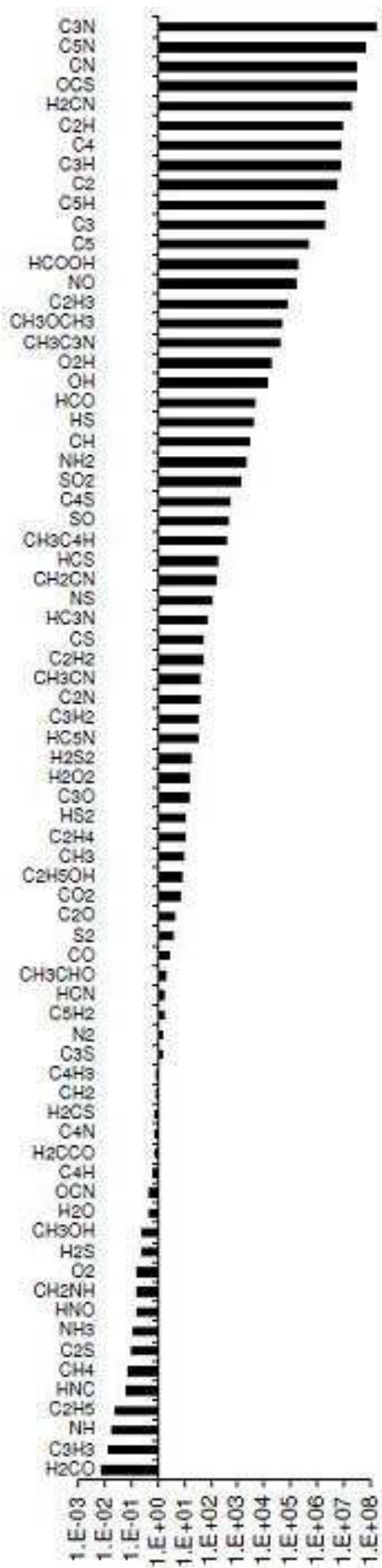}
	\caption{The normalized mantle-to-surface (nM/S) ratio for selected species, obtained from Table~\ref{table1} data.}
	\label{att-nMS}
\end{figure}

The calculation results are summarised in Tables~\ref{table1},~\ref{table2}, and~\ref{table3} and in Fig.~\ref{att-nMS}.

The overall results of the model are shown in Table~\ref{table1}. Results (Table~\ref{table2}) show that the mantle-to-surface (M/S) ratios for heavy elements are different from each other. This means that, for example, M/S of $\mathrm{C_{5}}$ and $\mathrm{O_{2}}$ molecules is not directly comparable, because the M/S ratio of carbon itself is higher than that of oxygen (cf. Table~\ref{table2}) and there is a larger abundance of carbon species overall in the mantle.  To be able to compare the mantle-to-surface ratios of different molecules, we calculate a weighed average elemental M/S ratio (\textit{WR}) for each individual specie according to

   \begin{equation}
   \label{res1}
WR_{m}=\frac{\sum^{i}_{j}(a\times RE)}{\sum^{i}_{j} a}.
   \end{equation}
\textit{a} (the index in a molecular formula) is the number of atoms of elements \textit{i} to \textit{j} in the molecule \textit{m}. \textit{RE} is the average M/S ratio for each element in the molecule \textit{m} (C, N, O, Na, Mg, Si, S or Fe) from \textit{i} through \textit{j}. Because hydrogen wanders rather freely between the surface and the mantle, it is not counted here. We name the relation

   \begin{equation}
   \label{res2}
\frac{(M/S)_{m}}{WR_{m}}
   \end{equation}

\textit{the normalized mantle-to-surface ratio} (nM/S). For species whose fractional abundance is equal in the mantle and on the surface, nM/S is unity. If the concentration in mantle is higher this value is higher than unity and vice versa. The nM/S for selected species are graphically shown in Fig.~\ref{att-nMS}.

The cosmic-ray-induced photodissociation yield in lower mantle layers should be rather low, because with yields larger than $10^{-3}$ in our simulations we obtain a mixture of compounds with water not the dominant molecule, which is inconsistent with the present knowledge about interstellar grain mantle composition. In this case the remaining oxygen is lost mostly to highly oxidized carbon-bearing species.

Hydrogen in its atomic, molecular, and chemically bound states is a special case because it is the only element capable of crossing the boundary between mantle and surface (Sect. ~\ref{Hdiff}). The data regarding ``solid'' phases of H are shown in Table~\ref{table3}.
As a standard comparison data for this paper (Table~\ref{table1}) we use the results given by Hasegawa \& Herbst (\cite{10}), Table 1, t=1.0E+8 yr. To properly compare the results we also calculated the molecular nM/S ratios for these data. Both works aim to investigate the solid phase chemistry in molecular clouds, and both have a 3-phase system (gas, surface, mantle). There are several important differences between the two models. Our model includes a reduced number of solid state species (see Sect.~\ref{consider}). The model of Hasegawa \& Herbst lacks cosmic-ray-induced photodissociation of solid state species (Sect.~\ref{CRPHdiss}), mantle-phase reactions (Sect.~\ref{reacs}) and hydrogen diffusion (Sect.~\ref{Hdiff}).

The main difference in the chemical modeling results is that we obtain more oxidized compounds, like $\mathrm{H_{2}O_{2}}$ in the mantle. They are mainly  $\mathrm{CO_{2}}$, HCOOH and less $\mathrm{CH_{4}}$ for carbon, $\mathrm{SO_{2}}$ for sulfur, more HCN, $\mathrm{N_{2}}$ and less $\mathrm{NH_{3}}$ for nitrogen. Generally one can say the improvements of our solid phase model diminishes the dominance of hydrogenated species and leads to a much greater chemical diversity with high abundances of fully or partially oxidized species. Another substantial difference in our model is the high concentration of atomic and other radical species, both on the surface and in the mantle. It can be attributed to the inclusion of photodissociation reactions for all solid species. Our model produces a much better overall fit to observations for gas phase species. It is because molecules dominant in the mantle phase are slowly returned into gas and partially reprocessed, and a total freeze-out for any species cannot occur.

\onecolumn
\begin{table}
\caption{Calculated fractional abundances of species in gas, surface, and mantle phases, and the normalized mantle-to-surface ratio. Included species are with both gas and solid phases only.}
\label{table1}
\centering
\begin{tabular}{lcccc|lcccc}
\hline\hline
Species&Gas&Surface&Mantle&nM/S&Species&Gas&Surface&Mantle&nM/S\\
\hline





C&4.56E-09&1.32E-17&2.15E-14&4.62E+01&OCS&2.99E-10&1.54E-16&1.58E-07&2.84E+07\\
N&1.52E-08&1.25E-14&2.59E-14&6.78E-02&SO2&3.09E-12&1.37E-10&5.99E-06&1.25E+03\\
O&8.83E-08&7.44E-17&2.35E-14&1.01E+01&CH3&1.48E-08&2.82E-14&9.05E-12&9.14E+00\\
NA&4.71E-09&8.23E-08&2.90E-06&1.00E+00&NH3&5.81E-08&8.34E-08&2.94E-07&1.15E-01\\
MG&1.18E-07&2.01E-06&6.78E-05&1.00E+00&SIH3&3.08E-09&1.44E-14&3.08E-07&9.71E+05\\
SI&1.79E-08&1.21E-14&3.05E-07&1.15E+06&C2H2&3.72E-09&1.08E-08&1.94E-05&5.10E+01\\
S&1.88E-08&1.52E-14&7.44E-07&1.15E+06&H2O2&1.08E-08&1.60E-07&7.18E-05&1.43E+01\\
FE&1.62E-07&1.81E-06&5.61E-05&1.00E+00&H2S2&5.71E-11&5.82E-10&4.47E-07&1.81E+01\\
CH&6.73E-09&7.95E-20&7.81E-15&2.79E+03&H2CN&9.53E-12&1.27E-20&7.59E-12&1.82E+07\\
NH&2.37E-08&3.32E-14&1.92E-14&1.89E-02&H2CO&7.70E-08&7.21E-06&1.69E-06&7.03E-03\\
OH&3.79E-07&3.92E-13&1.61E-07&1.31E+04&C3H&1.22E-09&2.27E-15&5.35E-07&6.72E+06\\
C2&1.54E-09&1.14E-15&2.26E-07&5.63E+06&H2CS&3.50E-09&8.21E-08&2.64E-06&8.29E-01\\
CN&9.49E-09&2.14E-15&2.00E-06&2.84E+07&C4&9.91E-11&1.30E-16&3.47E-08&7.57E+06\\
N2&2.32E-08&3.31E-07&1.42E-05&1.40E+00&C3N&2.31E-11&3.36E-17&2.93E-07&2.56E+08\\
CO&3.92E-07&1.96E-07&1.73E-05&2.66E+00&C3O&3.06E-11&7.57E-10&3.65E-07&1.41E+01\\
SIH&5.05E-10&1.24E-14&2.92E-07&1.07E+06&C3S&1.49E-10&1.45E-09&7.24E-08&1.35E+00\\
NO&8.11E-08&4.12E-13&1.85E-06&1.45E+05&CH4&5.68E-08&1.74E-06&4.68E-06&7.65E-02\\
O2&7.34E-08&5.25E-08&2.73E-07&1.66E-01&SIH4&6.29E-08&1.29E-06&3.68E-05&1.29E+00\\
HS&8.58E-09&3.61E-12&6.15E-07&4.02E+03&C2H3&1.60E-10&9.65E-15&2.24E-08&6.60E+04\\
SIC&6.74E-10&8.84E-09&2.33E-07&9.21E-01&CH2NH&1.43E-08&2.14E-07&1.15E-06&1.63E-01\\
SIO&1.33E-07&1.65E-06&2.67E-05&6.05E-01&C3H2&2.60E-09&5.56E-09&6.28E-06&3.22E+01\\
CS&2.61E-09&5.35E-10&1.06E-06&5.13E+01&CH2CN&9.68E-11&1.29E-09&6.70E-06&1.54E+02\\
NS&2.04E-09&1.28E-10&5.33E-07&1.14E+02&H2CCO&5.20E-10&6.48E-09&1.70E-07&7.75E-01\\
SO&1.81E-11&9.27E-11&1.46E-06&4.28E+02&HCOOH&1.07E-12&1.31E-11&7.17E-05&1.68E+05\\
SIS&1.77E-15&1.87E-14&1.88E-07&3.12E+05&C4H&1.84E-09&1.56E-08&3.66E-07&6.66E-01\\
S2&2.22E-09&2.32E-08&3.74E-06&3.81E+00&HC3N&3.88E-10&5.16E-09&1.28E-05&7.31E+01\\
CH2&6.19E-09&1.44E-15&4.31E-14&8.52E-01&C5&9.15E-10&3.91E-16&6.25E-09&4.55E+05\\
NH2&7.30E-08&9.41E-14&6.08E-09&2.11E+03&C4N&2.47E-10&2.54E-09&7.15E-08&8.21E-01\\
H2O&2.84E-07&1.70E-05&2.42E-04&4.53E-01&C4S&3.17E-13&2.86E-12&5.18E-08&4.94E+02\\
SIH2&2.82E-10&1.25E-14&3.05E-07&1.10E+06&C2H4&8.64E-09&1.03E-09&3.73E-07&1.03E+01\\
C2H&6.91E-09&5.95E-15&1.72E-06&8.23E+06&CH3OH&2.58E-10&3.41E-09&3.05E-08&2.69E-01\\
O2H&5.08E-13&4.45E-14&2.63E-08&1.88E+04&C3H3&3.76E-09&1.20E-07&5.88E-08&1.40E-02\\
HS2&1.10E-10&1.14E-09&5.38E-07&1.11E+01&CH3CN&4.06E-11&4.71E-10&5.88E-07&3.71E+01\\
HCN&3.86E-08&9.45E-07&5.33E-05&1.71E+00&C5H&2.08E-10&6.81E-16&4.28E-08&1.79E+06\\
HNC&2.05E-08&3.07E-07&7.01E-07&6.94E-02&C5N&2.56E-11&1.10E-17&2.46E-08&6.48E+07\\
HCO&8.30E-09&4.53E-13&6.64E-08&4.41E+03&C2H5&4.76E-09&4.10E-07&3.77E-07&2.62E-02\\
HCS&5.96E-10&1.70E-15&1.08E-11&1.64E+02&C4H3&1.77E-09&2.07E-08&6.64E-07&9.11E-01\\
HNO&1.44E-08&1.44E-06&6.90E-06&1.55E-01&CH3CHO&1.26E-09&1.54E-08&9.53E-07&1.83E+00\\
H2S&8.06E-09&5.49E-07&6.25E-06&2.68E-01&C5H2&6.09E-10&2.30E-08&1.26E-06&1.55E+00\\
C3&1.59E-09&8.85E-16&5.29E-08&1.70E+06&HC5N&4.17E-11&6.21E-10&6.87E-07&3.22E+01\\
C2N&2.16E-10&2.24E-09&2.66E-06&3.53E+01&CH3C3N&1.72E-14&1.72E-13&2.41E-07&4.08E+04\\
C2O&2.06E-10&2.12E-09&3.28E-07&4.57E+00&CH3OCH3&2.38E-15&2.79E-14&3.92E-08&4.15E+04\\
C2S&5.84E-10&6.07E-09&2.35E-08&1.03E-01&C2H5OH&4.69E-13&5.16E-12&1.38E-09&7.87E+00\\
OCN&4.80E-08&4.75E-07&7.81E-06&5.08E-01&CH3C4H&3.93E-13&3.94E-12&5.15E-08&3.72E+02\\
CO2&3.89E-08&6.92E-07&1.64E-04&7.26E+00&&&&&\\

\hline
\end{tabular}
\end{table}
\twocolumn

\begin{table}
\caption{Total elemental fractional abundances in surface and mantle phases, calculated from Table~\ref{table1} data.}
\label{table2}
	\centering
		\begin{tabular}{cccc}
\hline\hline
Element&Surface&Mantle&Mantle-to-surface\\
\hline
C&1.35E-05&4.73E-04&3.51E+01\\
N&4.14E-06&1.27E-04&3.07E+01\\
O&2.98E-05&9.35E-04&3.14E+01\\
Na&8.23E-08&2.90E-06&3.53E+01\\
Mg&2.01E-06&6.78E-05&3.38E+01\\
Si&2.95E-06&6.50E-05&2.20E+01\\
S&6.89E-07&2.93E-05&4.24E+01\\
Fe&1.81E-06&5.61E-05&3.10E+01\\
\hline
		\end{tabular}
\end{table}

\begin{table}
\caption{Fractional abundances for hydrogen species in surface and mantle phases.} \label{table3}
	\centering
		\begin{tabular}{cccc}
\hline\hline
Species&Surface&Mantle&Mantle-to-surface\\
\hline
$\mathrm{H_{2}}$&1.17E-05&2.34E-07&2.00E-02\\
H&1.25E-18&2.17E-20&1.73E-02\\
H (chemically bound)&6.84E-05&1.12E-03&1.64E+01\\
\hline
		\end{tabular}
\end{table}

The main results of our investigation are:
\begin{enumerate}
	\item Important molecules on the surface are $\mathrm{H_{2}O}$, $\mathrm{H_{2}CO}$, $\mathrm{CH_{4}}$. Those in the mantle are $\mathrm{H_{2}O}$, $\mathrm{CO_{2}}$, HCOOH, $\mathrm{H_{2}O_{2}}$ (Table ~\ref{table1}). The most abundant molecule among both the surface and mantle species is water. However, the molecule containing most of the oxygen in the mantle is $\mathrm{CO_{2}}$. In our opinion this represents the general shift from highly to poorly hydrogenated species by the photochemical processing of the mantle.
	\item The nM/S ratio for highly hydrogenated species is usually rather low, in the range 0.1 - 0.01 (Fig.~\ref{att-nMS}). The most important species here are $\mathrm{CH_{4}}$ and other saturated carbon molecules. Methane has a high concentration on the surface, which significantly decreases in the mantle, releasing huge amounts of carbon now available for various oxidized and chain-like compounds.
	\item	The highest nM/S is seen for cyanopolyyne related molecules, showing that mantle processing efficiently transforms saturated hydrocarbons to these species. The related HNC is consumed in the mantle, and the HCN/HNC gas abundance ratio of 1.9 is adequate for dark cores. Remarkable are the high nM/S values for oxidized sulfur species (OCS, SO, $\mathrm{SO_{2}}$), although these molecules have a very limited reaction set. The abundance of S oxides in the mantle exceeds that of $\mathrm{H_{2}S}$. We expect that hydrogen depletion in the mantle should essentially explain the high abundance of OCS and sulfur oxides in hot core regions noted by many observers (e.g. Mookerjea et al. \cite{34}, Hatchell et al. \cite{35}). We predict that 3-phase models with a more complete reaction network will produce OCS and other sulfur species in an amount more consistent with the observations.
	\item According to calculations, the hydrogen depletion in the mantle has different effects on the abundances of carbon-oxygen compounds. Molecules HCOOH and HCO have large nM/S, for $\mathrm{C_{3}O}$, $\mathrm{C_{2}O}$, $\mathrm{CH_{3}CHO}$ it is mediocre, and for $\mathrm{CH_{3}OH}$, $\mathrm{H_{2}CO}$ nM/S is less than unity. With a limited degree of certainty one can conclude that weakly hydrogenated C-O species are those favored by the conditions inside grain mantles, which is confirmed by observations of hot cores (e.g. Mookerjea et al. \cite{34}). Molecules $\mathrm{CH_{3}OCH_{3}}$, $\mathrm{C_{2}H_{5}OH}$ and $\mathrm{H_{2}CCO}$ have a very deficient chemical reaction set and are not considered here.
	\item An increased radical content in the mantle. Species like C, O, CH, S and others (notably excluding nitrogen species) show nM/S ratios moderately higher than 1. The model-based explanation is that only a limited reaction network is available and that not all species produced by the dissociation of CR induced photons are able to readily react and generate stable molecules. However, the real radical content depends on photodissociation yields and on the exact conditions in the mantle (see Sect.~\ref{credibl}).
\end{enumerate}

\section{Discussion}
\label{discussion}
  There is some experimental evidence available in the literature that backs up the results. Ferrante et al. (\cite{32}) and Garozzo et al. (\cite{46}) show experimentally that OCS is readily formed in laboratory-simulated interstellar conditions, while our model shows OCS to be a molecule with a very high nM/S ratio. However, we note again that OCS has a very limited reaction set. Several sources (e.g. Weaver et al. \cite{33} and references therein) note high abundances of organic molecules with H, C and O in regions of grain mantle disruption or evaporation, indicating that these species are formed on grains. This agrees with our model results, where most of these species have nM/S higher (much higher for some molecules) than unity (Fig.~\ref{att-nMS}).
  
Our model shows that the hydrogen diffusion outwards from the mantle is much more significant than the inward diffusion. The species initially included in the mantle are hydrogen-rich due to the surface reactions. Photoprocessing releases the chemically bound hydrogen, it diffuses, escapes to the surface and, eventually, the gas phase. According to the model, this process is not overcome by the hydrogen diffusing into the mantle from outside.

Although the hydrogen diffusion coefficient in amorphous, dirty ice is most probably smaller than the one we used (in hexagonal ice), the model is insensitive to the exact value of $D$ over several orders of magnitude. This is the case as long as the time of the diffusion is much shorter than the heavy molecule residing time in the mantle. Also, the model is rather insensitive to the exact value of inward diffusion probability $P_{diff,SM}$, if it is kept lower (for example, by a factor of 2/3) than the outward $P_{diff,MS}$=0.5. The solid phase model results are essentially determined by the reasonable estimate (see Sect.~\ref{Hdiff}) that $P_{diff,MS}>P_{diff,SM}$. With a degree of certainty one can say that the total direction of hydrogen diffusion at 10\textit{K} (in, immobile, or out) is determined by the structure of the mantle, not by the exact diffusion speed.

The hydrogen atoms released in mantle photodissociation are very slow to diffuse and are very reactive. However, among other molecules, there always form $\mathrm{H_{2}}$ molecules that are generally much less reactive and able to easily diffuse and leave the mantle. While hydrogen is depleted, methane and other highly saturated hydrocarbons transform to cyanopolyynes and other hydrogen-poor carbon chain compounds (Fig.~\ref{att-nMS}). Water slowly loses hydrogen and the remaining oxygen forms compounds with other metals (in our model it is mostly $\mathrm{CO_{2}}$).

The most serious counter-argument produced by our model is the overabundance of the $\mathrm{CO_{2}}$ molecule in calculation results. However, like many others, this important species is badly affected by a poor reaction set, its only sink reaction is the slow dissociation by cosmic-ray-induced photons. This is also the exact reason why chains with six and more carbon atoms were excluded from our calculations. The concentration of overproduced molecules would be reduced by interaction with radicals (including atomic hydrogen) in the mantle, meanwhile reducing the amount of radicals themselves. In the reaction set by Hasegawa, Herbst \& Leung (\cite{18}) and Hasegawa \& Herbst (\cite{11}) most important radicals that lack solid phase reactions have the general formula $\mathrm{\textit{M}H_{\textit{x}}}$ where \textit{M} is C, N, O, S or Si. They are generated by photodissociation in our model. The reactions between any complicated species and a variety of radicals are desirable for future mantle chemical models.

The differences in elemental abundance within mantle and surface (Table~\ref{table2}) can be clearly accounted for selective desorption effects by direct cosmic-ray-heating and by $\mathrm{H_{2}}$ formation. Carbon is accumulated because it tends to form chain-like heavy molecules, while sulfur molecules generally have high binding energies and also do not easily desorb.

The hydrogen-poor conditions within the grain mantles may promote a much more colorful chemistry than is usually thought about interstellar matter. In the model it can be seen best with the C-O compounds. Reasoning from the calculation results we expect that the sub-surface mantle conditions would favor an assortment of (1) rather complex and (2) oxidized species. If we attribute these properties to the heavier elements, we can conclude that a significant proportion of Si, P, and S should be highly oxidized and in molecules forming many bonds. Strong inorganic or semi-organic acids, their salts and esters with the organic alcohols, peroxy acids, etc., can be some of the compounds whose formation can be permitted by the lack of hydrogen in the mantle. To estimate the abundance of these compounds, one needs a greatly expanded solid-phase reaction set.

Last but not least, we offer our explanation to the observable ``lack'' of sulfur molecules in dark clouds (see e.g. Goicoechea et al. \cite{39}, Wakelam et al. \cite{40}, Palumbo et al. \cite{44}). The calculated large abundance and n/MS of sulfur in highest oxidation state, namely $\mathrm{SO_{2}}$, in our model output allow us to suppose that the conditions within the mantles are suitable for many oxidized sulfur compounds that can essentially be described as derivatives of the sulfuric and sulfurous acids. These would include esters, sulfinic and sulfonic acids, sulfoxides, sulfones, peroxysulfuric acids, sulfamides, sulfimides, sulfamic acids, their derivatives and acid salts. Plenty of these molecules are known to be stable at 273K and should be durable in the mantles at 10K if they are able to form. Organic species include unsaturated thiols, thioethers and thioketones. Other species would be carbon disulfide $\mathrm{CS_{2}}$ and those containing an S-S bond, already mentioned by Garozzo et al. (\cite{46}). Our idea is that a significant amount (say, some 40\%) of sulfur is dissipated over a great variety of minor species (including many simpler radicals) and thus is hardly observable. An important assumption to explain why the derivatives are not observable in hot cores, is that when these species evaporate, they are reduced by hydrogen and dissociated in fragments by radiation. That is, these molecules lack gas production pathways. We suspect that a higher abundance of oxidized sulfur (mostly SO, $\mathrm{SO_{2}}$, CS) bearing species should be observable in the middle evolution stages of a hot core. Indeed, these observation results are presented, for example, by Mookerjea et al. (\cite{34}), Wakelam et al. (\cite{42}), Wakelam et al. (\cite{40}), Chandler et al. (\cite{41}), Jim\'{e}nez-Serra et al. (\cite{43}) with abundances of $\mathrm{SO_{2}}$ and, especially, SO typically larger than those of $\mathrm{H_{2}S}$ and comparable to OCS. A less well pronounced but similar effect can also be expected for the phosphorus family of compounds, as this element also is known to have a rich chemistry. One may argue that the ``disappeared'' sulfur in dark cores resides on grains in the form of oxides. This is however not confirmed by observations, e.g. Palumbo et al. (\cite{44}). It is well known that sulfur oxides tend to combine with water, which is thought to be the dominant molecule on interstellar grains. In our opinion mantle sulfur chemistry can be expected to lead to the various acidic derivatives, besides other more conventional compounds. Some investigations of the sulfur chemistry (Palumbo et al. \cite{44}, Ferrante et al. \cite{32}) insist that the most important sulfur molecule known in dark cores OCS forms in water-deficient ices, so sulfur acid derivatives perhaps would be what one can expect from water-rich ices.

\section{Conclusions}
\label{conclusions}

\begin{enumerate}
\item According to our model, which includes several basic assumptions (see Sect.~\ref{model}), it is highly possible that chemical processes below the outer surface of interstellar grain mantles play an important role in the chemistry of dark molecular cloud cores.
\item An important factor that should be further investigated is the hydrogen diffusion through the grain mantles. This, combined with the continued dissociation of molecules by cosmic ray induced photons, leads to an overall outward flux of hydrogen from the mantle. According to our model results, the more dense the grain mantle, the more efficient is the outward diffusion of hydrogen. A more diverse, H-poor chemistry is encouraged below the surface, explaining the abundance of at least some species observed in dark molecular clouds and hot molecular cores.
\item A model based on the concept of pore surface reactions can at least partially describe the transformations occurring within the icy mantle.
\item Fe nuclei of cosmic-rays can be a cause of physical alteration of the mantle structure, but other possibilities, like light cosmic-ray particles, CR induced UV photons, and slow thermal diffusion may provide alteration with generally similar chemical consequences.
\item A combination of outer and inner mantle surface chemistry is able to produce a wide set of mantle species. It may ultimately lead to more accurate calculations of the composition of molecular clouds.
\item Further research is required to clarify many factors that are only approximately estimated. These include photodissociation and desorption yields, H and $\mathrm{H_{2}}$ diffusion rate in amorphous dirty ices, the thickness of the mantle, the properties and number of the pores, reaction mechanism inside the grain mantles, effectiveness of selective desorption mechanisms, etc.
\item A chemically active and hydrogen poor environment in the mantle may explain the difficulties of observing sulfur in dark molecular cores. In grain mantles the rich chemistry of sulfur permits the formation of many various S molecules (mostly oxoacid derivatives) low on hydrogen, most of them with abundances too low to be observed. The large abundance of sulfur oxides in hot star-forming cores may be a direct consequence of these compounds being ejected into the gas phase.
\end{enumerate}


\onecolumn
\begin{longtable}{ll|ll|ll|ll}
\caption{\label{gasconc} Calculated gas phase fractional abundances of molecules.}\\
\hline\hline
Species&Abundance&Species&Abundance&Species&Abundance&Species&Abundance\\
\hline
\endfirsthead
\caption{continued.}\\
\endhead
\endfoot

H&8.39E-06&C7&1.21E-10&H2+&5.86E-14&N2+&4.80E-18\\
HE&1.67E-01&C7+&1.71E-18&H2C3H+&2.84E-13&N2O&0.00E+00\\
O&7.40E-08&C7H&9.39E-12&H2C5N+&5.31E-15&N2O+&0.00E+00\\
N&1.59E-08&C7H+&6.19E-16&H2C7N+&1.51E-16&NA+&5.24E-09\\
C&4.19E-09&C7H2&1.83E-11&H2C9N+&7.54E-18&NCCN&7.11E-09\\
MG&8.94E-08&C7H2+&2.21E-16&H2CCC&6.80E-10&NCCNCH3+&8.76E-20\\
S&1.56E-08&C7H3+&2.65E-16&H2CCCC&1.14E-10&NCCNH+&3.86E-14\\
SI&1.72E-08&C7H4&1.70E-14&H2CCO&4.00E-10&NH&2.41E-08\\
FE&1.24E-07&C7H4+&2.05E-18&H2CN&9.89E-12&NH+&4.85E-16\\
NA&3.60E-09&C7H5+&7.15E-19&H2CO&5.15E-08&NH2&7.00E-08\\
C10&5.15E-14&C7N&2.81E-12&H2CO+&5.87E-13&NH2+&5.84E-15\\
C10+&1.19E-18&C7N+&1.58E-24&H2CS&2.39E-09&NH2CN&1.48E-07\\
C2&1.39E-09&C8&7.76E-12&H2CS+&2.65E-14&NH2CNH+&8.63E-13\\
C2+&1.13E-17&C8+&1.51E-17&H2NC+&1.26E-13&NH3&5.97E-08\\
C2H&6.23E-09&C8H&6.24E-12&H2NO+&4.07E-13&NH3+&3.59E-12\\
C2H+&3.32E-17&C8H+&1.27E-19&H2O&2.05E-07&NH4+&2.26E-12\\
C2H2&3.37E-09&C8H2&1.66E-12&H2O+&1.11E-14&NO&6.82E-08\\
C2H2+&5.79E-15&C8H2+&1.71E-16&H2S&5.58E-09&NO+&5.38E-13\\
C2H3&1.54E-10&C8H3+&1.71E-17&H2S+&1.79E-13&NO2&5.80E-10\\
C2H3+&7.16E-14&C8H4+&6.59E-19&H2S2&3.22E-11&NO2+&1.87E-20\\
C2H3CN+&2.27E-23&C8H5+&1.02E-22&H2S2+&2.50E-15&NS&1.65E-09\\
C2H4&6.88E-09&C9&5.89E-12&H2SIO&2.20E-09&NS+&5.68E-14\\
C2H4+&1.70E-14&C9+&4.55E-19&H2SIO+&1.38E-14&O+&9.33E-16\\
C2H5&4.68E-09&C9H&4.63E-13&H3+&1.21E-09&O2&5.26E-08\\
C2H5+&3.06E-14&C9H+&3.29E-17&H3C3O+&2.42E-19&O2+&6.64E-14\\
C2H5OH&3.09E-13&C9H2&1.03E-12&H3C3OH3+&1.69E-20&O2H&3.30E-13\\
C2H5OH+&3.64E-18&C9H2+&1.14E-17&H3C3OH4+&3.82E-19&O2H+&2.60E-16\\
C2H5OH2+&6.39E-17&C9H3+&1.52E-17&H3C5N+&4.74E-21&OCN&2.51E-08\\
C2H7+&5.48E-21&C9H4+&2.95E-20&H3C7N+&7.73E-20&OCN+&2.66E-19\\
C2N&1.69E-10&C9H5+&4.50E-22&H3C9N+&6.50E-21&OCS&2.00E-10\\
C2N+&4.57E-14&C9N&1.28E-13&H3CO+&2.86E-12&OCS+&2.81E-16\\
C2N2+&1.64E-21&C9N+&8.39E-26&H3CS+&5.99E-14&OH&3.18E-07\\
C2NH+&5.92E-15&CH&5.98E-09&H3O+&2.74E-11&OH+&2.68E-15\\
C2O&1.41E-10&CH+&1.29E-16&H3S+&2.88E-13&S+&5.64E-11\\
C2O+&5.54E-16&CH2&6.24E-09&H3S2+&5.26E-15&S2&1.23E-09\\
C2S&3.88E-10&CH2+&4.24E-16&H3SIO+&7.84E-14&S2+&8.83E-15\\
C2S+&9.39E-19&CH2CHCN&1.86E-19&H4C3N+&2.85E-23&SI+&1.96E-10\\
C3&1.24E-09&CH2CN&5.92E-11&H5C2O2+&2.26E-20&SIC&5.77E-10\\
C3+&7.92E-17&CH2CN+&1.44E-15&HC2O+&1.51E-15&SIC+&1.05E-17\\
C3H&9.02E-10&H2CCO+&4.19E-16&HC2S+&1.01E-14&SIC2&1.38E-09\\
C3H+&6.74E-16&CH2NH&1.09E-08&HC3N&2.60E-10&SIC2+&8.79E-18\\
C3H2&1.99E-09&CH2NH2+&4.97E-15&HC3N+&5.17E-16&SIC2H&5.96E-10\\
C3H2+&5.77E-14&CH3&1.48E-08&HC3NH+&9.73E-15&SIC2H+&1.07E-13\\
C3H2O+&2.41E-21&CH3+&9.61E-13&HC3O+&1.15E-15&SIC2H2&1.34E-09\\
C3H3&2.51E-09&CH3C3N&9.30E-15&HC3S+&3.78E-15&SIC2H2+&2.80E-14\\
C3H3+&6.44E-14&CH3C3NH+&1.73E-18&HC4N+&1.44E-19&SIC2H3+&4.94E-14\\
C3H5+&1.05E-16&CH3C4H&2.76E-13&HC4S+&1.80E-17&SIC3&2.09E-10\\
C3H6+&2.30E-25&CH3C4H+&2.68E-18&HC5N&2.19E-11&SIC3+&6.35E-15\\
C3H7+&2.42E-22&CH3C5N&2.56E-16&HC5N+&8.46E-19&SIC3H&7.54E-11\\
C3N&1.55E-11&CH3C5NH+&1.16E-19&HC7N&2.20E-12&SIC3H+&1.72E-14\\
C3N+&1.78E-18&CH3C7N&2.53E-17&HC7N+&1.56E-17&SIC3H2+&9.68E-15\\
C3O&2.98E-11&CH3C7NH+&1.17E-20&HC9N&1.04E-13&SIC4&1.09E-14\\
C3O+&4.77E-20&CH3CCH&5.03E-10&HC9N+&7.17E-19&SIC4+&3.92E-15\\
C3S&8.99E-11&CH3CCH+&6.64E-15&HCN&2.97E-08&SIC4H+&1.47E-18\\
C3S+&1.05E-15&CH3CH3&7.85E-13&HCN+&4.31E-16&SICH2&8.53E-09\\
C4&7.28E-11&CH3CH3+&2.52E-14&HCNH+&5.90E-12&SICH2+&1.23E-13\\
C4+&3.27E-17&CH3CHO&9.17E-10&HCO&6.88E-09&SICH3&8.18E-13\\
C4H&1.35E-09&CH3CHO+&1.19E-14&HCO+&4.55E-12&SICH3+&7.51E-14\\
C4H+&9.11E-17&CH3CHOH+&1.28E-14&HCO2+&3.52E-14&SICH4+&2.41E-17\\
C4H2+&9.43E-14&CH3CN&2.45E-11&HCOOCH3&1.20E-16&SIH&5.85E-10\\
C4H3&1.18E-09&CH3CN+&2.95E-18&HCOOH&6.37E-13&SIH+&5.38E-13\\
C4H3+&1.57E-15&CH3CNH+&1.29E-14&HCOOH+&5.42E-26&SIH2&3.53E-10\\
C4H4+&9.70E-15&CH3CO+&1.09E-14&HCOOH2+&6.76E-17&SIH2+&9.78E-15\\
C4H5+&1.39E-16&CH3COCH3&1.04E-15&HCS&5.03E-10&SIH3&3.12E-09\\
C4H7+&6.76E-20&CH3CS+&2.98E-16&HCS+&5.12E-14&SIH3+&6.38E-14\\
C4N&1.77E-10&CH3OCH3&1.35E-15&HCSI&6.95E-09&SIH4&5.22E-08\\
C4N+&4.72E-18&CH3OCH3+&1.13E-20&HCSI+&3.00E-17&SIH4+&2.99E-17\\
C4S&1.87E-13&CH3OCH4+&1.73E-19&HE+&6.38E-10&SIH5+&4.84E-13\\
C4S+&1.10E-15&CH3OH&1.84E-10&HEH+&1.02E-15&SIN&1.82E-09\\
C5&6.22E-10&CH3OH+&4.53E-16&HN2+&4.84E-14&SIN+&2.08E-13\\
C5+&6.18E-18&CH3OH2+&1.04E-15&HN2O+&0.00E+00&SINC+&4.95E-14\\
C5H&1.34E-10&CH4&5.04E-08&HNC&1.60E-08&SINH2+&2.63E-13\\
C5H+&8.34E-15&CH4+&1.92E-16&HNO&1.17E-08&SIO&9.20E-08\\
C5H2&3.96E-10&CH4N+&1.85E-15&HNO+&2.88E-13&SIO+&1.78E-15\\
C5H2+&1.26E-14&CH5+&1.37E-12&HNS+&8.05E-14&SIO2&2.35E-08\\
C5H3+&6.90E-15&CN&8.15E-09&HNSI&1.02E-08&SIOH+&4.97E-12\\
C5N&1.42E-11&CN+&2.58E-16&HNSI+&2.77E-17&SIS&1.71E-15\\
C5N+&2.13E-23&CNC+&7.26E-14&HOC+&9.55E-16&SIS+&2.19E-16\\
C6&1.18E-10&CO&2.95E-07&HS&6.52E-09&SO&1.20E-11\\
C6+&2.35E-17&CO+&3.45E-16&HS+&3.09E-13&SO+&2.35E-13\\
C6H&8.46E-11&CO2&2.10E-08&HS2&6.12E-11&SO2&1.65E-12\\
C6H+&8.34E-16&CO2+&2.03E-18&HS2+&7.71E-15&SO2+&2.73E-19\\
C6H2&3.26E-11&COOCH4+&1.47E-21&HSIO2+&1.83E-13&C+&5.39E-11\\
C6H2+&1.86E-15&CS&1.84E-09&HSIS+&7.94E-19&C5H5+&8.08E-18\\
C6H3+&3.20E-16&CS+&5.39E-17&HSO+&5.94E-16&C6H6&1.07E-12\\
C6H4+&1.05E-17&FE+&1.96E-07&HSO2+&5.49E-17&MG+&2.67E-08\\
C6H5+&1.46E-19&H+&2.21E-10&N+&1.22E-13&H2O2&5.94E-09\\
C6H6+&1.49E-16&H2&8.32E-01&N2&1.96E-08&E&2.33E-07\\
C6H7+&1.19E-17&&&&&&\\

\hline
\end{longtable}
\twocolumn

\end{document}